\documentclass[twocolumn,secnumarabic,amssymb, nobibnotes, aps, prd]{revtex4-1}

\usepackage{graphicx}
\usepackage[latin1]{inputenc}

\begin{document}

\newcommand{\nx}{\textrm}

\title{Magnetoresistance in a High Mobility Two-Dimensional Electron Gas}

\author{L. Bockhorn$^1$}
\author{P. Barthold$^1$}
\author{D. Schuh$^2$}
\author{W. Wegscheider$^2$\footnote{present address: Laboratorium für Festkörperphysik, ETH Zürich, Schafmattstr. 16, 8093 Zürich, Switzerland}}
\author{R. J. Haug$^1$}
\affiliation{$^1$Institut f\"ur Festk\"orperphysik, Leibniz Universit\"at Hannover, Appelstr. 2, 30167 Hannover, Germany\\
$^2$ Institut für Experimentelle und Angewandte Physik, Universität Regensburg, Universit\"atsstrasse 31, 93053 Regensburg, Germany}
\date{\today}

\begin{abstract}
In a high mobility two-dimensional electron gas (2DEG) in a GaAs$\:$/$\:$Al$_{0.3}$Ga$_{0.7}$As quantum well we observe a strong magnetoresistance. In lowering the electron density the magnetoresistance gets more pronounced and reaches values of more than 300$\%$. We observe that the huge magnetoresistance vanishes for increasing the temperature. An additional density dependent factor is introduced to be able to fit the parabolic magnetoresistance to the electron-electron interaction correction.
\end{abstract}


\maketitle
Since the first observation of the fractional quantum Hall effect (FQHE)\cite{Tsui1982, Laughlin1983} the quality and the mobility of the two-dimensional electron gas (2DEG) increased by more than two orders of magnitude. The increased mobility has allowed not only the observation of the FQHE at many different filling factors and smaller magnetic fields but also many new effects. So, microwave-induced oscillations were observed, which are up to now not fully understood\cite{Zudov2001a, Mani2002, Zudov2003}. In weak magnetic fields the increased mobility enabled also the observation of phonon-induced resistance oscillations, which are caused by inelastic scattering between electrons and three dimensional acoustic phonons\cite{Zudov2001, Raichev2009}. The period of phonon-induced oscillations is tunable by an additional dc electric field\cite{Yang2002,Zhang2008}. Also a new type of QHE was enabled in high mobility 2DEGs, the re-entrant integer quantum Hall effect (RIQHE)\cite{Eisenstein2001, Eisenstein}. In the regime of the RIQHE the longitudinal resistance between integer filling factors decreases to zero suggesting fractional filling factors, but the corresponding Hall plateaus are quantized at integer values.

Here we will present the observation of a huge magnetoresistance in a high mobility 2DEG which depends strongly on electron density and temperature.

\begin{figure}
   \centering
   \includegraphics{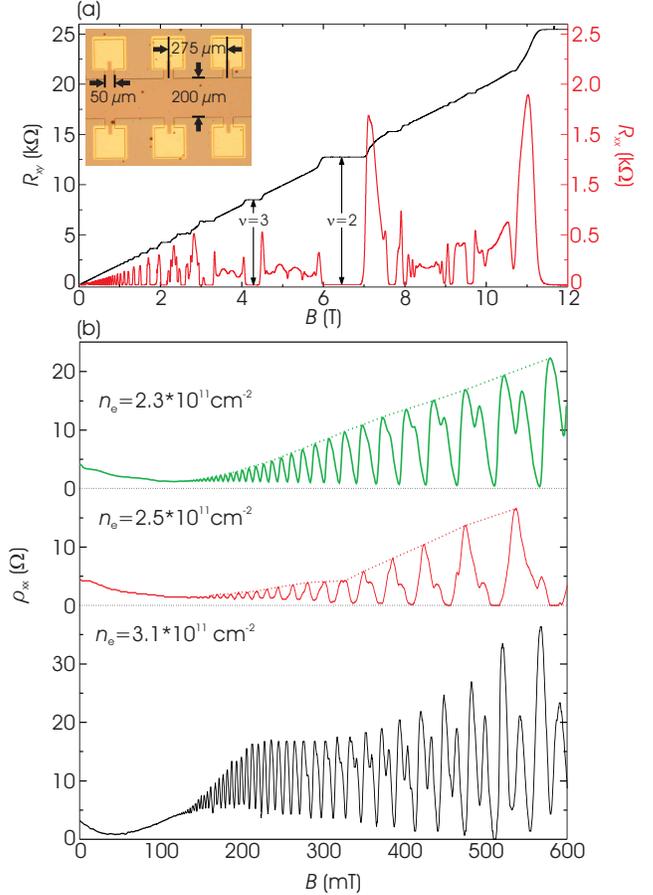}
   \caption{\label{Fig1} (a) Hall resistance$\;R_{xy}$ (black curve) and longitudinal resistance$\;R_{xx}$ (red curve) vs. magnetic field$\;$B at 45$\;$mK. The electron density is $n_{e}\approx3.1\cdot10^{11}$cm$^{-2}$ and the mobility is $\mu\approx11.9\cdot10^{6}$cm$^{2}$/Vs. The inset shows a part of the Hall bar. (b) The longitudinal resistivity$\;\rho_{xx}$ vs. magnetic field$\;$B for different electron densities in the range 0$\;$mT and 600$\;$mT.}
\end{figure}

Our samples were cleaved from a wafer of a high-mobility GaAs/Al$_{0.3}$Ga$_{0.7}$As quantum well grown by molecular-beam epitaxy. The quantum well has a width of 30$\;$nm and is Si-doped from both sides. The 2DEG is located 150$\;$nm beneath the surface and has an electron density of $n_{e}\approx3.1\cdot10^{11}\;$cm$^{-2}$ and a mobility of $\mu\approx11.9\cdot10^{6}\;$cm$^{2}$/Vs in the dark. The specimens are Hall bars with a total length of 1.2$\;\mu$m, a width of $w=200\;\mu$m and a potential probe spacing of $l=275\;\mu$m (see Fig.$\;\ref{Fig1}$(a)). The Hall bars were defined by photolithography and wet etching. Different ungated and gated samples were used for the magnetotransport measurements. In case of the gated sample there is an additional layer of 600$\:$nm PMMA between the Hall bar and the metallic topgate to avoid leakage current. We apply topgate voltages up to -6$\;$V to manipulate the electron density. Our measurements were performed in a dilution refrigerator with a base temperature of 20$\;$mK. The measurements were carried out by using low-frequency (13$\;$Hz) lock-in technique.

Figure$\;\ref{Fig1}$(a) shows the longitudinal resistance$\;R_{xx}$ and the Hall resistance$\;R_{xy}$ vs. magnetic field$\;$B to demonstrate the quality of our samples. A series of different fractional quantum Hall states appears for filling factor $\nu<2$. We observe also the filling factor $\nu=5/2$. Over the range $4<\nu<6$ the longitudinal resistance decreases to zero between integer filling factors with Hall plateaus being quantized at integer values of $h/4e^{2}$, $h/5e^{2}$ and $h/6e^{2}$. The here observed phenomenon is the RIQHE.

In Fig.$\;\ref{Fig1}$(b) the longitudinal resistance$\;R_{xx}$ vs. magnetic field$\;$B is shown for different electron densities$\;n_e$ in the range 0$\;$mT to 600$\;$mT. We observe a peak at zero magnetic field. The strong negative magnetoresistance crosses over to a positive magnetoresistance at about 40$\;$mT for the highest electron concentration. The Shubnikov-de$\;$Haas$\;$(SdH) oscillations show for $n_{e}=3.1\cdot10^{11}\;$cm$^{-2}$ a beating effect, which is also observed till $n_{e}=2.5\cdot10^{11}\;$cm$^{-2}$. For $n_{e}=2.3\cdot10^{11}\;$cm$^{-2}$ no beating effect is observable. So, the beating effect disappears by decreasing the electron density. This beating effect for $2.5\cdot10^{11}\;$cm$^{-2}<n_{e}<3.14\cdot10^{11}\;$cm$^{-2}$ is attributed to the existence of two 2D subbands. The occupation of the second subband occurs above $n_{e}=2.5\cdot10^{11}\;$cm$^{-2}$. This low value is attributed to the double Si-doped quantum well and is lower than previously reported values (see e.g. \cite{Hamilton1995}). The SdH oscillations start at 110$\;$mT. From this onset of the SdH oscillations we can deduce the density inhomogeneity of our samples\cite{Syed2004} and we find an inhomogeneity of less than 2$\%$.

\begin{figure}
   \centering
   \includegraphics{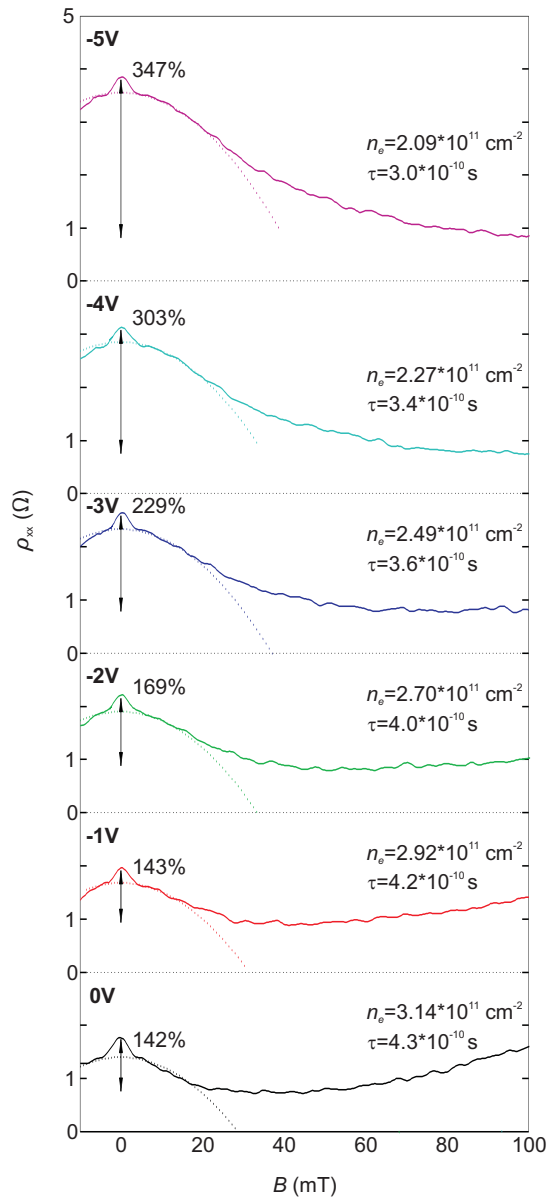}
   \caption{\label{Fig3}The longitudinal resistivity$\;\rho_{xx}$ vs. magentic field$\;$B for different topgate voltages. The different curves have been shifted vertically for clarity.}
\end{figure}
In Fig.$\;$\ref{Fig3} the longitudinal resistivity$\;R_{xx}$ vs. magnetic field$\;$B is shown for different topgate voltages. For each topgate voltage we achieve a huge magnetoresistance. The longitudinal resistivity decreases almost to a similar background by applying a magnetic field. The difference between this value and the peak is indicated by the percentage (see Fig.$\;$\ref{Fig3}). The magnetoresistance increases by decreasing the electron density$\;n_{e}$ and the difference is reaching a value of about 350$\;\%$. The width of the huge peak in the magnetoresistance increases also with decreasing electron density. The longitudinal magnetoresistance becomes nearly bell-shaped for lower electron densities. At zero magnetic field a small peak is observed on top of the bell-shaped magnetoresistance. This small peak appears for all topgate voltages.

\begin{figure}
   \centering
   \includegraphics{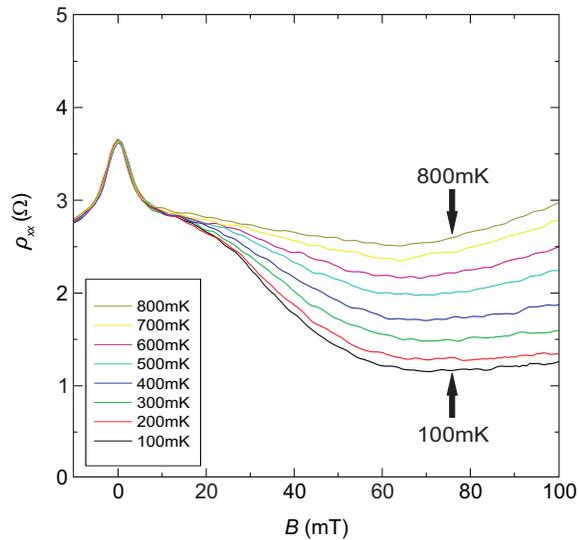}
   \caption{\label{Fig2} The longitudinal resistivity$\;\rho_{xx}$ vs. magnetic field$\;$B for several temperatures, ranging between 100$\;$mK and 800$\;$mK. The peak at zero magneticfield has nearly the same value for all temperatures.}
\end{figure}
For gate voltages being larger than -2$\;$V the negative magnetoresistance crosses over to a positive magnetoresistance at about 40$\;$mT, while for lower topgate voltages the magnetoresistance does not show such a positive magnetoresistance. Therefore the crossover at 40$\;$mT is attributed to the additional 2D subband being occupied for a carrier concentration of $n_{e}=3.14\cdot10^{11}\;$cm$^{-2}$. For $n_{e}<2.5\cdot10^{11}\;$cm$^{-2}$ we do not have any hint to the occupation of the second 2D subband, but we see the huge negative magnetoresistance. Therefore the astonishing behavior of the huge magnetoresistance is not caused by the interaction between different 2D subbands. 
 
The magnetoresistance depends not only strongly on the electron density but also on temperature. In Fig.$\;$\ref{Fig2} the longitudinal resistivity$\;\rho_{xx}$ vs. magnetic field$\;$B is shown for a gate voltage of -4$\;$V for several temperatures ranging between 100$\;$mK and 800$\;$mK. The SdH oscillations start outside of the shown magnetic field range. 
Similar to the topgate voltage dependent measurements we observe the huge magnetoresistance for the lowest temperature, 100$\;$mK, and on top of the huge magnetoresistance the small peak at zero magnetic field.

The negative magnetoresistance decreases by increasing the temperature. Meanwhile the small peak at zero magnetic field is left unchanged by increasing the temperature. The temperature independence at $B=0\;$T is a sign for the absence of weak localization in our sample. In contrast to the huge magnetoresistance which depends strongly on temperature, the peak at zero magnetic field is temperature independent. Since the mean free path of our sample is about 113$\;\mu$m and the Hall bar dimensions are in the range of the mean free path, one can attribute the observed effect to the influence of ballistic transport. The peak at zero magnetic field is then given by scattering at the edges of the geometry of our Hall bars in the ballistic transport regime, comparable to the effects observed in the so-called quenching of the Hall effect\cite{Roukes1987,Thornton1989}.

From the above observations we know that the astonishing behavior of the huge magnetoresistance is neither caused by weak localization nor the interaction between different 2D subbands. The observed effect has to be related to the high mobility of the 2DEG, the corresponding mean free path and interaction effects. 

To compare our measurements with other in the literature mentioned effects we examine the electron interaction correction to the conductivity$\;\delta\sigma^{ee}_{xx}(T)\;$ \cite{Paalanen1983, Gornyi2003, Li2003}. The negative magnetoresistance is in accordance with Li$\;$et$\;$al.\cite{Li2003} expressed by
\begin{equation}
	\rho_{xx}=\frac{1}{\sigma_{0}}+\frac{1}{\sigma^{2}_{0}}\left(\mu^{2}B^{2}\right)(\delta\sigma^{ee}_{xx}(T))^{-1}
	\label{eq:1}
\end{equation}
where $\sigma_{0}$ is the Drude conductivity. This expression includes temperature dependence and a parabolic magnetoresistance produced by long-range potential scattering. Since our sample shows $\tau>>\tau_{q}$, where $\tau_{q}$ is the quantum time determined from the magnitude of SdH oscillations to the order of magnitude of $10^{-13}\;$sec, we can conclude that the scattering is dominated by long-range scatterers. The huge parabolic magnetoresistance is analyzed in the range of strong fields $\omega_{c}\tau>1$ which satisfy the theoretical approximations. On the basis of the transport scattering time of our samples and the equation
\begin{equation}
	T\geq\frac{\hbar}{k_{B}}\cdot\frac{1}{\tau}
	\label{eq:2}
\end{equation}
our measurements take place in the ballistic regime\cite{Gornyi2003,Gornyi2004}. The electron interaction induced correction to the conductivity considering the influence of the ballistic transport is then expressed by
\begin{equation}
(\delta\sigma^{ee}_{xx}(T))^{-1}=-\frac{e^{2}}{\pi\;h}\;c_{0}\alpha\sqrt{\frac{\hbar}{T\tau\;k_{B}}}.
	\label{eq:3}
\end{equation}
$c_{0}$ is according to reference\cite{Paalanen1983, Gornyi2003, Li2003} constant and has a value of about $c_{0}=0.276$. We introduce an additional factor $\alpha$ to fit our resistivities.
In Fig.$\;$\ref{Fig3} we fit the resistivity as a $B^{2}$ dependence up to 20$\;$mT for different topgate voltages and obtain $(\delta\sigma^{ee}_{xx}(T))^{-1}$ using eq.$\;$(\ref{eq:1}). The electron interaction induced correction to the conductivity only fits for our measurements if $\alpha>1$. In Fig.$\;\ref{Fig3}$ one sees that the maximum of the achieved negative parabola increases by increasing the electron density, while the curvature of the parabola is unchanged. From this observation we achieve $\alpha$ which depends strongly on the electron density and varies between 30 and 150. 

\begin{figure}
   \centering
   \includegraphics{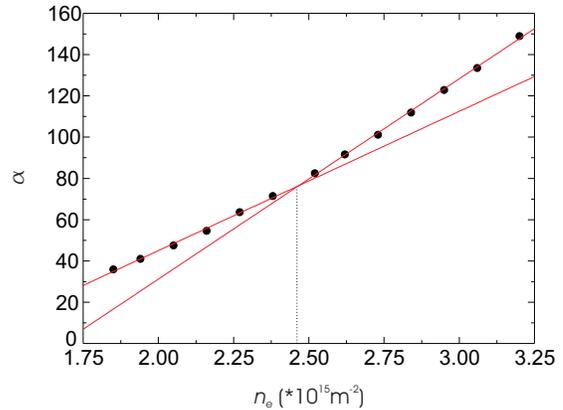}
   \caption{\label{Fig4} The factor $\alpha$ vs. electron density$\;n_e$. The dashed line marks the crossover from one occupied 2D subband to two occupied 2D subbands.}
\end{figure}
Figure$\;$\ref{Fig4} shows the dependence of the factor$\;\alpha$ on the electron density. With increasing electron density $\;\alpha$ increases also. Between $2.4\cdot10^{11}\;$cm$^{-2}$ and $2.5\cdot10^{11}\;$cm$^{-2}$ a kink is clearly observed in the increase of the factor$\;\alpha$ with the density. A linear approximation for the $\alpha$ values on each side of the kink results in a crossover at $n_{e}=2.46\cdot10^{11}\;$cm$^{-2}$. Above we have shown that a second 2D subband is occupied for $n_{e}>2.5\cdot10^{11}\;$cm$^{-2}$. The kink marks the crossover from one occupied 2D subband to two occupied 2D subbands. Thus, the factor$\;\alpha$ clearly depends not only on the total electron density but also on the scattering properties of the sample which are changed for the occupation of a second 2D subband\cite{Stoermer1982}.

Although we can fit the huge magnetoresistance to a parabolic magnetic field dependence as predicted for the electron interaction correction to the conductivity according to references\cite{Paalanen1983, Gornyi2003, Li2003} we had to introduce this factor$\;\alpha$ to describe the huge magnetoresistance. From the electron interaction induced correction$\;$eq.$\;$(\ref{eq:3}) the parabola curvature is expected to depend on $T^{-1/2}$. For a given gate voltage we observe that the parabola curvatures changes by increasing the temperature, while the parabola maximum is unchanged (see Fig.$\;$\ref{Fig2}). We observe a hint towards the expected temperature dependence of $T^{-1/2}$ for the lowest temperatures of our experimental data. Above 200$\;$mK the temperature dependence is more complex. A possible origin for the discrepancy between theory and experiment could be that the influence of the density fluctuation for high mobility 2DEG is not correctly described. In these high mobility samples the very small, but finite density variation across the sample induces an additional long range potential, up to now not treated in theory. A more sophisticated theoretical model of the electron interaction correction to the conductivity seems to be needed to describe these high mobility samples.

In conclusion, we observed for different gated and ungated samples a huge magnetoresistance which depends strongly on the electron density and the temperature. The huge parabolic magnetoresistance is fitted by the interaction correction to the conductivity in the situation of a long-range fluctuation potential and in the regime of ballistic transport and a discrepancy to theory is observed. 

We would like to thank F.$\;$Hohls for useful discussions and for help with the experiments. This work was supported by QUEST.


\end{document}